\documentclass[twoside,12pt]{article} 
\usepackage{ppnp}      
\usepackage{mod}
\usepackage{graphicx,subfigure}
\usepackage{amsmath,amssymb,amsfonts}
\usepackage{german}
\begin{document}
\newcommand{\engl}{\selectlanguage{english}}
\newcommand{\germ}{\selectlanguage{german}}
\engl
\title{Hadron structure at small momentum transfer.}
\author{Thomas Walcher\\
\large Institut f\"ur Kernphysik\\ Johannes-Gutenberg-Universit\"at Mainz}
\maketitle

\begin{abstract}
Giving three examples, the form factors of the nucleon, the polarisability 
of the charged pion and the interference of the $S_{11}(1535)$ with the 
$D_{13}(1520)$ excitation of the nucleon in the $\eta\,p$-decay channel, 
it is argued that the hadron structure at low momentum transfer is highly
significant for studying QCD.  
\end{abstract} 

\section{Introduction: Significance of low $Q^2$} 

There is little agreement about the most significant questions and the final  
aim of the field of hadron physics. One formulation due to an old NSAC 
definition is: ``Understand how hadrons are made up of quarks and gluons''. 
However, from this definition it is not clear whether one has a study, or 
even a check, of QCD proper in mind or one wants just to understand bound 
systems of quarks and gluons with ``QCD inspired'' models. Since QCD is not 
yet solvable for low $Q^2$, the region where the hadrons in nature reside, 
the skeptics turn away and join more promising fields.  

Some who stay, put their hope into lattice QCD which, however, is so far good 
for large quark masses only. The other limit, valid for vanishing quark masses,
is chiral dynamics. It reflects directly QCD symmetries and vacua. However, 
this approximation fails frequently to describe the data, in particular at 
finite momentum transfers, even if they are small.

An alternative view of hadron physics is to seek the aim in the understanding 
of the effective degrees of freedom, emerging from the many body system of 
quarks and gluons. This means one takes QCD for the given right theory of 
strong interactions and is interested in understanding the peculiar facets of 
hadrons as confinement, spontaneous symmetry breaking and the strong coupling 
constant at small momentum transfer. 

The characteristic feature of many body systems is that they produce emerging
structure which are equivalent to effective degrees of freedom. These degrees
of freedom are e.g. constituent quarks, di-quarks, pions or vector mesons 
allowing an approximate description of many observables. If one establishes
the salient features of such degrees of freedom by significant experiments one 
may have finally the chance to extrapolate between the mentioned low and large 
quark mass approximations of QCD.    

The most burning question in hadron physics is therefore: What are the most
significant experiments for studying the effective degree of freedom. Many
believe that old and new structure functions like spin-structure function,
transversity or generalised parton distributions are the right experiments. 
It is the conviction of this author that the experiments at low $Q^2$,
i.e. $Q^2 \lessapprox 5\,$Gev$^2$, are at least as significant for such 
studies. These experiments study gross properties of hadrons as form factors, 
polarisabilities, resonances, and sum rules which can be directly confronted
with models based on effective degrees of freedom. 

In this article three examples of significant observables are given: 
the nucleon form factors, the polarisability of the charged pion, and 
the $S_{11}(1535)$ and the $D_{13}(1520)$ excitations of the nucleon in 
the $\eta\,p$-decay channel. All three examples hint to deviations 
from our standard understanding of hadron structure. The significance is 
just in the deviations between theory and experiment since they offer the 
chance to learn something new. 

For a more detailed account of the arguments in this introduction see
ref.~\cite{Drechsel:2007aa}.

\section{Form factor of the nucleons, revisited}

\begin{figure}[h]
\begin{center}
\includegraphics[width=6cm,angle=-90]{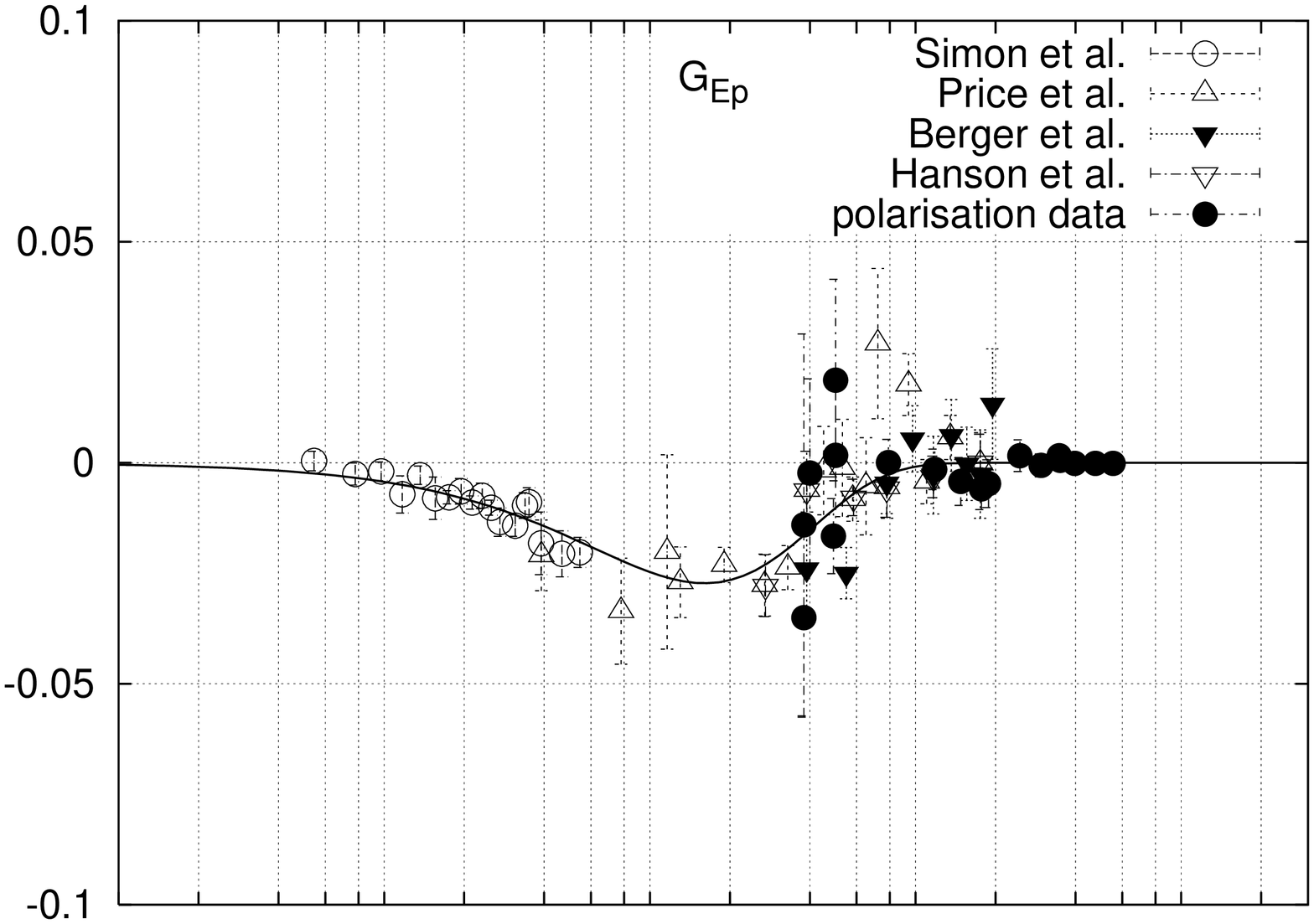}%
\includegraphics[width=6cm,angle=-90]{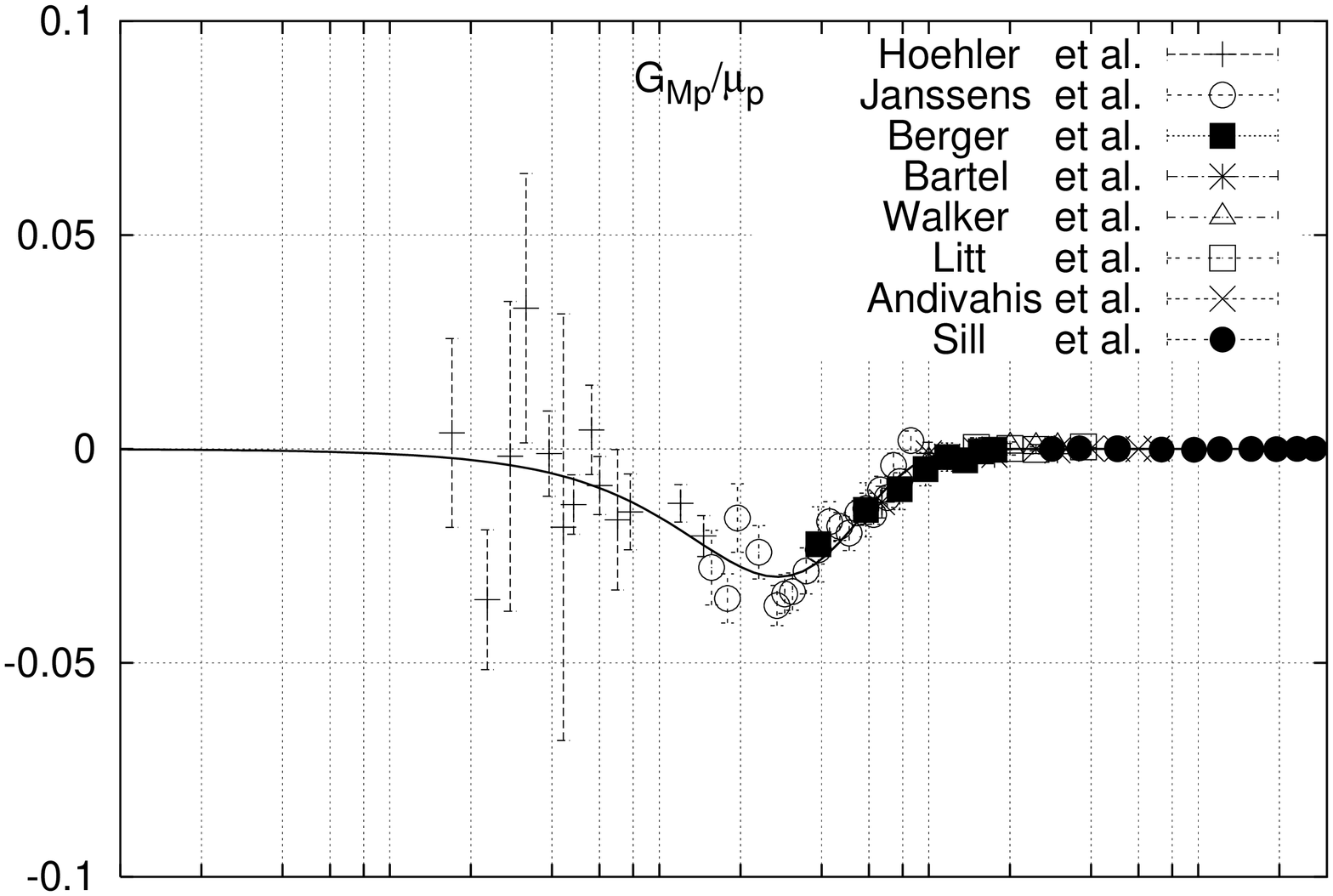}

\includegraphics[width=6cm,angle=-90]{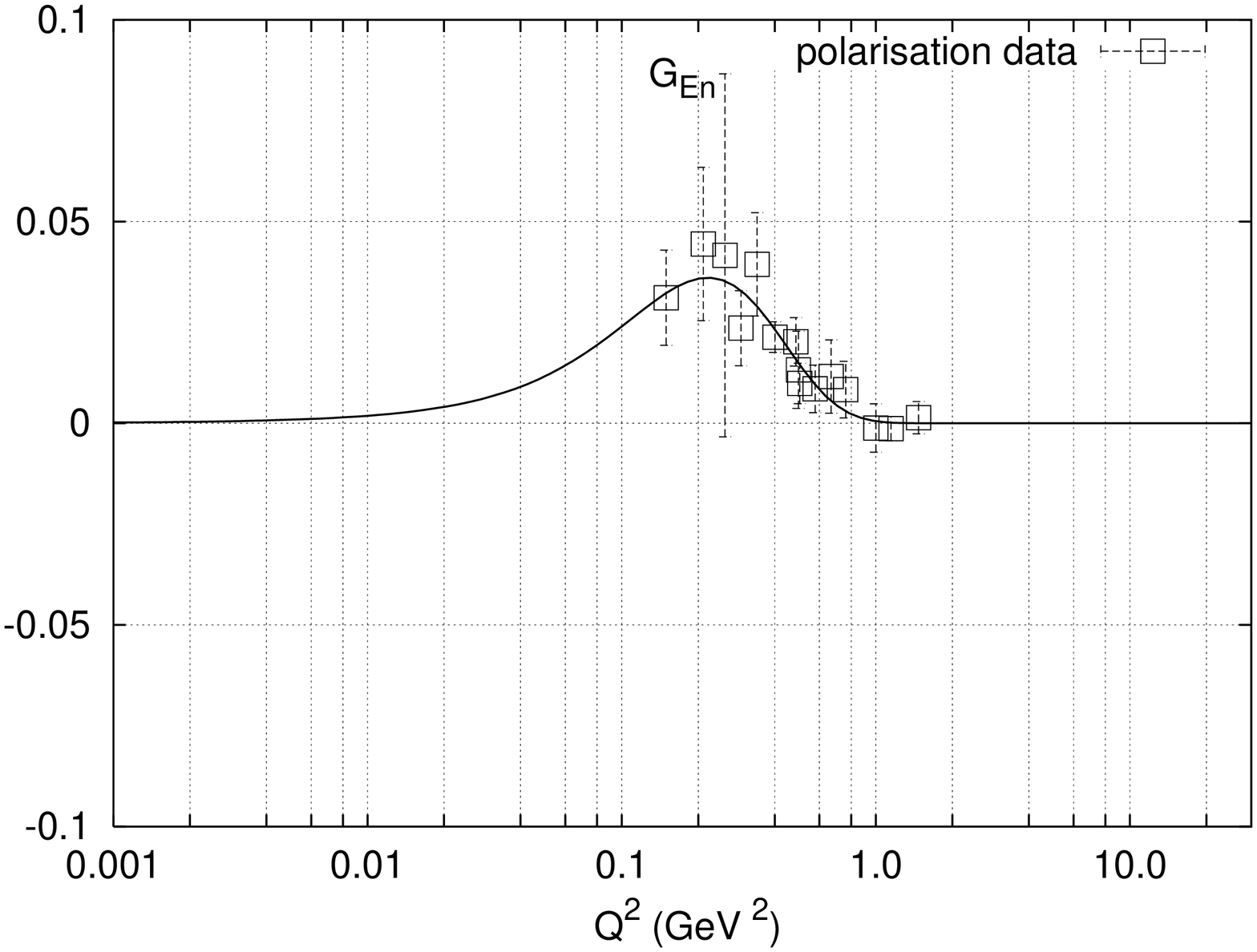}%
\includegraphics[width=6cm,angle=-90]{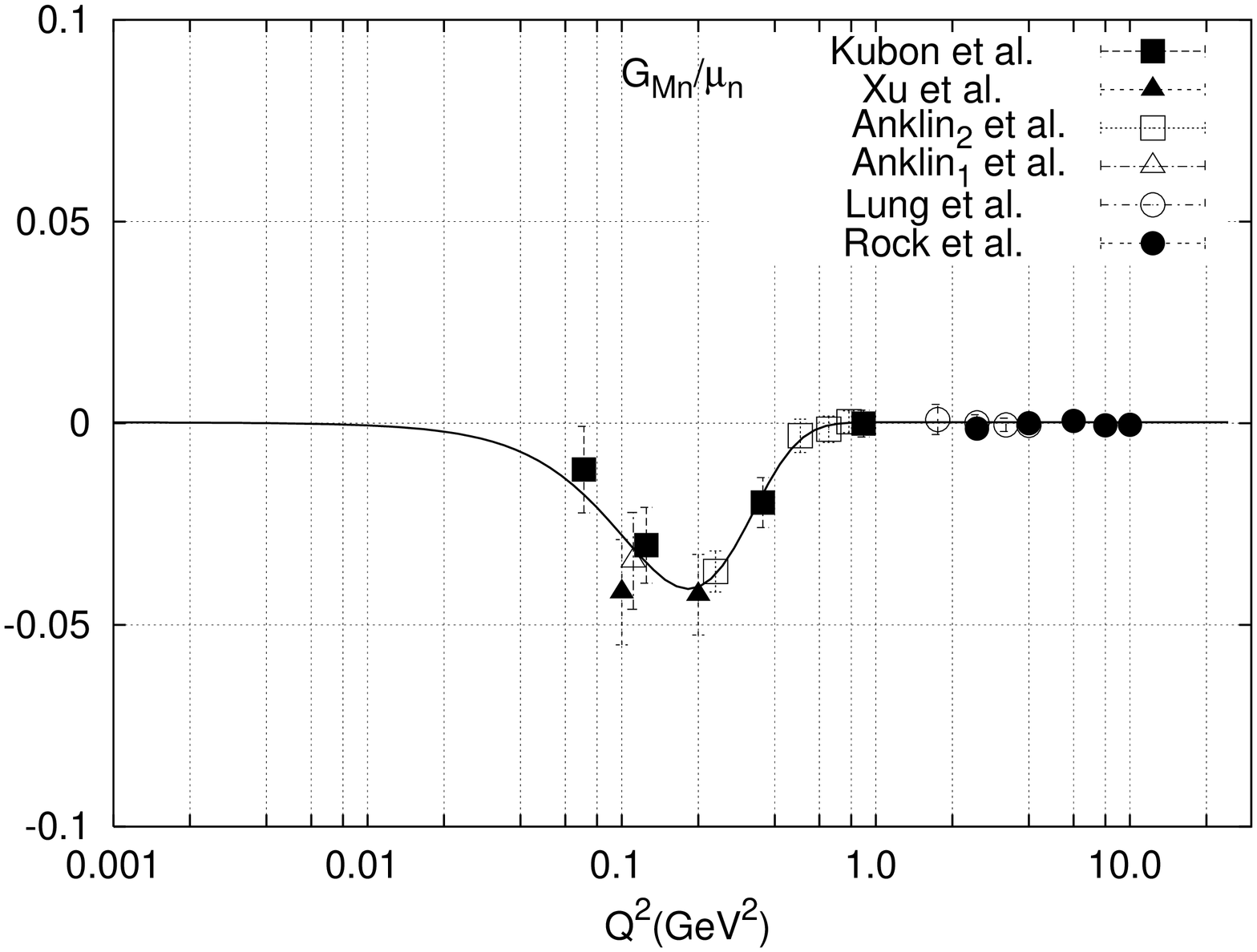} 
\end{center}
\caption{The electric and magnetic form factors of the proton and neutron
  after subtraction of a smooth bulk contribution with short wave length 
  in r-space. (From ref.~\cite{Friedrich:2003ab})}
\label{fig:ff_all}
\end{figure}

Our knowledge of form factors has greatly improved over the last decade due to 
new experimental opportunities, mainly at Jlab, Bates-MIT and MAMI. The most
remarkable features are probably the dramatic deviation of the electric form
factor of the proton from the dipole form and the much improved accuracy for
the electric form factor of the neutron. Particularly at low momentum
transfers the possible appearance of bumps and dips with large wave length 
in r-space has stirred some controversy about their theoretical
interpretation. Before we comment this controversy, we want to show the world
data of the space like form factors again. Figure~\ref{fig:ff_all} shows these
data together with a fit of a somewhat simple minded model. The model assumes
a core of constituent quarks ($m_{\text{cq}} \approx 300\,$MeV) distributed in 
Q-space as the sum of two dipole forms and surrounded by a pion cloud 
represented by a Gaussian bump. An alternative model uses a nonrelativistic 
p-state wave function for the pion cloud. The fits of these two models in 
ref.~\cite{Friedrich:2003ab} imply clearly that the splitting of the charge
distribution into a ``bare nucleon'' and a ``pion cloud'' is highly model 
dependent. Therefore, these models must not be taken to seriously as a
realistic physical picture of the relativistic system nucleon. Rather they are
a convenient way to describe the data by a smooth curve allowing to 
Fourier-transform it for a visualisation of the charge distribution in the 
Breit frame. 

\begin{figure}[h]
\begin{center}
\subfigure[The world data for the electric form factor of the neutron $G_E^n$.]{
\begin{minipage}{8.5cm}
\includegraphics[width=8.5cm,angle=0]{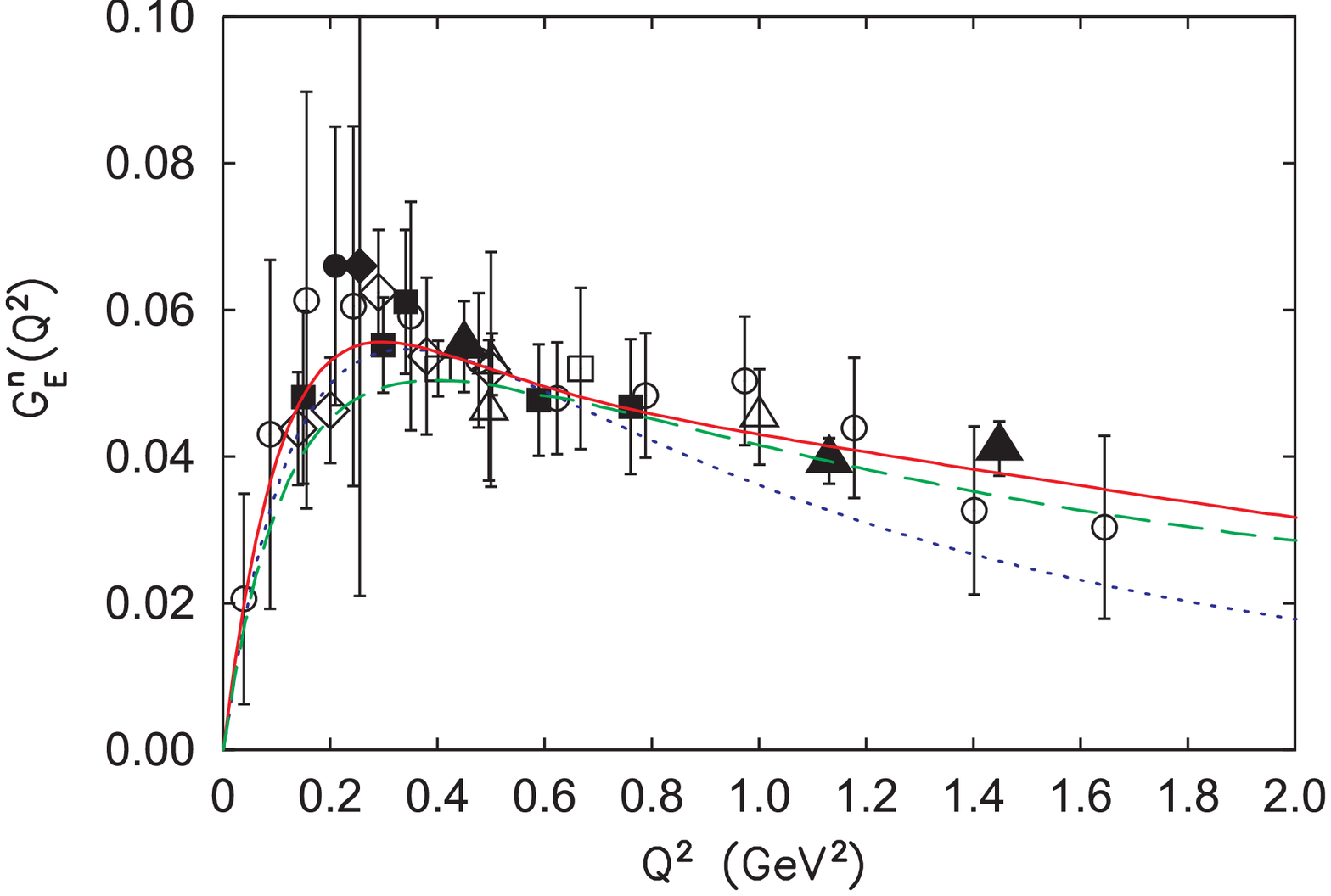}
\label{fig:GE_n}
\end{minipage}}%
\hspace{5mm}
\subfigure[Visualisation of the charge distribution $4\pi\,r^2\,\rho(r)$ in
           the Breit frame.]{
\begin{minipage}{8.5cm}
\centering
\includegraphics[width=8.5cm,angle=0]{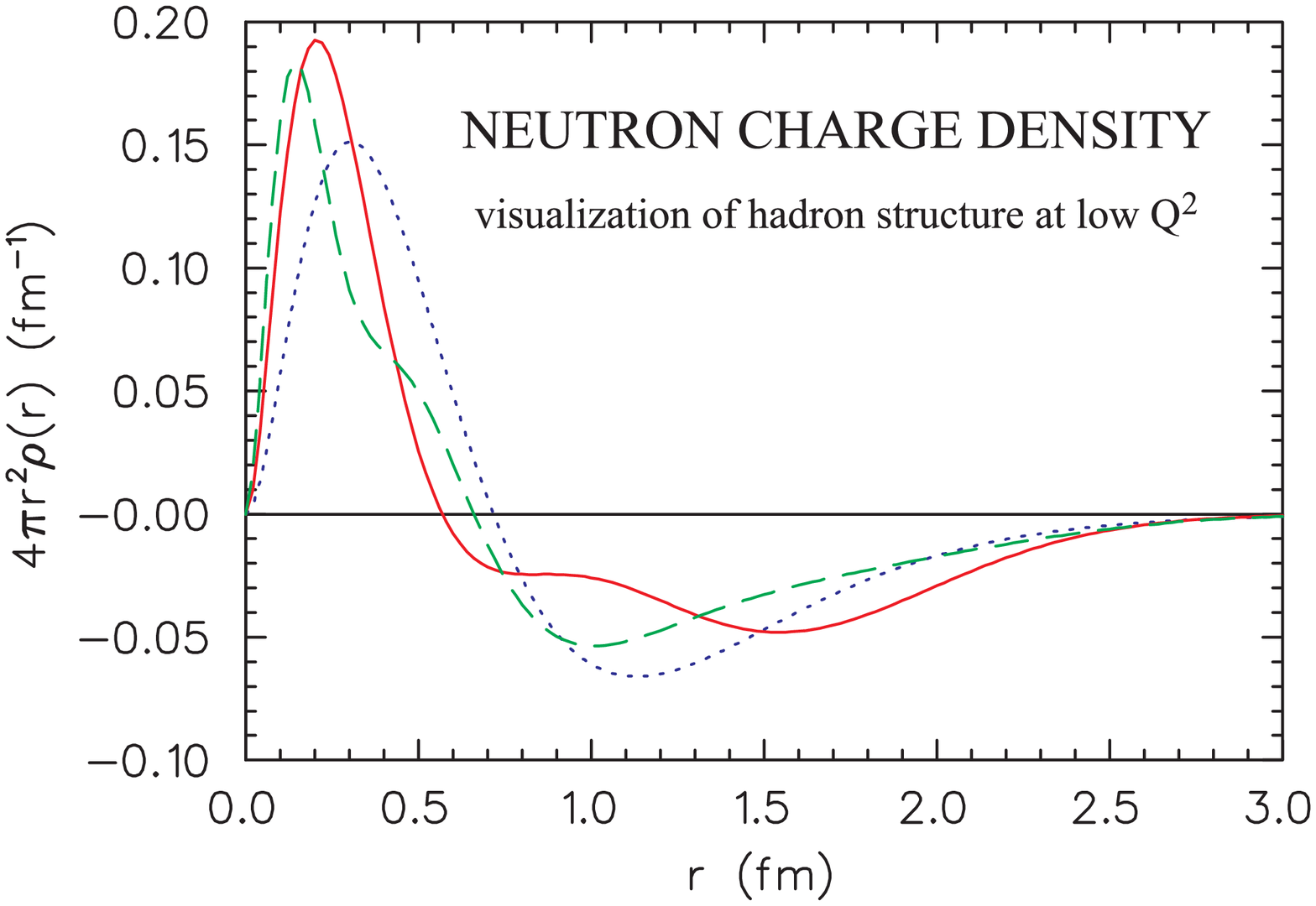}
\label{fig:rho_n}
\end{minipage}}
\caption{The dotted curve represents the old Galster parametrisation, the
  dashed curve the dispersion relation calculation of Belushkin, Hammer and
  \germ Mei"sner \engl \cite{Belushkin:2006qa}, and the solid curve is a new 
  fit of the phenomenological model of ref.~\cite{Friedrich:2003ab}.}
\end{center}
\end{figure}

The particularly interesting charge distribution of the neutron is shown
in Fig.~\ref{fig:rho_n} which is based on a new fit to the world date for 
the electric form factor $G_{En}$ depicted in Fig.~\ref{fig:GE_n}. A good fit 
of the existing data indicates charge at large radii. It is natural to identify
this charge with an effective pion, the lightest colour neutral hadron which
may reach out beyond the confinement radius of approximately 1~fm. 

The idea behind this phenomenological analysis has been harshly criticised
recently \cite{Meissner:2007zz} since the analysis based on the fundamental 
dispersion-relations is in clear disagreement (see Fig.~\ref{fig:rho_n}). 
However, the explanation of this disagreement as a purely statistical problem
in ref.~\cite{Belushkin:2006qa} is not valid. The dispersion relation fits of 
all nucleon form factors in the space like and time like region gives a 
$\chi^2/\text{dof} \approx 1.8$ for about 200 degrees of freedom (dof).  
The statistical probability for such a large $\chi^2$ $P(\chi^2/\text{dof} 
\approx 1.8, \text{dof} \approx 200)$ is smaller than $10^{-10}$. Therefore, 
the 1-$\sigma$ bands in \cite{Belushkin:2006qa} determined by adding 1 to the 
absolute $\chi^2$ are meaningless. There exist more discrepancies between the 
dispersion relations and the data as discussed in some detail in 
ref.~\cite{Drechsel:2007aa}. These disagreements between a fundamental theory 
and several different experiments are intriguing and, therefore, their further 
study is very significant.      

\section{Pion polarisability, revisited}
Another significant observable for which stringent theoretical predictions
from Chiral Perturbation Theory (ChPTh) exist \cite{Gasser:2006cd} is the 
polarisability of the charged pion. On the experimental side the situation 
was until recently completely confused. The values determined ranged from 2 
to  $20\unit$ \cite{Ahrens:2004mg}. Therefore, the new determination at MAMI 
\cite{Ahrens:2004mg} promised hope to clarify the issue. However, the
new value of $(\a-\b)_{\pp} = (11.6\pm 1.5_{stat}\pm 3.0_{syst}\pm 
0.5_{model})\unit$ is in contradiction to the theoretical value of 
ref.~\cite{Gasser:2006cd} $(\alpha+\beta)_{\pi^+} = (0.16 \pm 0.1) \unit$
and $(\alpha-\beta)_{\pi^+} = (5.7~ \pm 1.0) \unit$. Such a deviation meant a
dramatic problem for ChPTh. But, since the deviation has a 2-$\sigma$ 
significance only and a not well controlled model dependence, this 
determination was taken with scepticism. This scepticism seemed to be 
confirmed by a new determination of the COMPASS collaboration at CERN 
using the Primakoff scattering of a pion from nuclei. This collaboration 
got a preliminary value of $(\a-\b)_{\pp} = (5.0\pm 3.4_{stat}\pm 1.2_{syst}
\pm ?_{model})\unit$ \cite{Colantoni:2007nn} much more in line with ChPTh. 
However, this value is at clear variance with the older determination at 
Serpukov based on the same Primakoff scattering \cite{Antipov:1983df} 
yielding $(\a-\b)_{\pp} = (13.6 \pm 2.8_{tot} \pm ?_{model}) \unit$. 
Also the recent reanalysis of $\gamma \gamma \rightarrow \pi^+ \pi^-$ 
by Fil'kov and Kasheharov results in the large value 
$(\a-\b)_{\pp} = (13.0^{+2.6}_{-1.9})\unit$
\cite{Filkov:2006ho} consistent with the one at Serpukov and MAMI.

It is, therefore, interesting to trace possible differences between the
Serpukov and the COMPASS experiment. It turned out in several talks given by
members of the COMPASS collaboration (\cite{Filkov:2007ha} and this author) 
that one decisive difference is in the cut on the maximal $Q^2_{\text{max}}$ 
of the differential cross section of the pion scattering. This cut has to be 
chosen small enough to guarantee that the contribution due to the strong 
interaction below the Coulomb peak is negligible. Whereas Serpukov used the 
very small limit of $Q^2_{\text{max}} \approx 6 \times 10^{-4}$\,GeV$^2$ the 
COMPASS collaboration allowed $Q^2_{\text{max}} \approx 5 \times 
10^{-3}$\,GeV$^2$. If one cuts the preliminary COMPASS data at the same low 
value as Serpukov the COMPASS value increases by about a factor of 2 and 
agreement between the two Primakoff experiments is established.

It is not difficult to understand this behaviour. The differential cross
section of the Primakoff scattering can be described by two amplitudes: the
Coulomb amplitude and a diffractive strong interaction amplitude. In order to
get at the Compton scattering effect one selects the $\gamma \pi^-$ channel 
by means of the detectors. Both amplitudes contain contributions to this final 
state and the selected statistical ensembles will interfere. Here enters a
frequent misunderstanding. The diffraction at the high energies given here 
will produce many unobserved particles and can be expanded in terms of Feynman
diagrams with different final states. Some believe that such ``diagrams'' 
would not interfere because they can be distinguished ``in principle''. 
However, such an idea is in conflict with quantum scattering theory which 
requests that one has to sum over all unobserved channels coherently 
(see e.g. \cite{Newton:1966de}). In many practical cases of the application 
of quantum mechanics the interference terms cancel or vanish because of 
orthogonality, but here they do not.

\begin{figure}[h]
\begin{center}
\includegraphics[height=5cm]{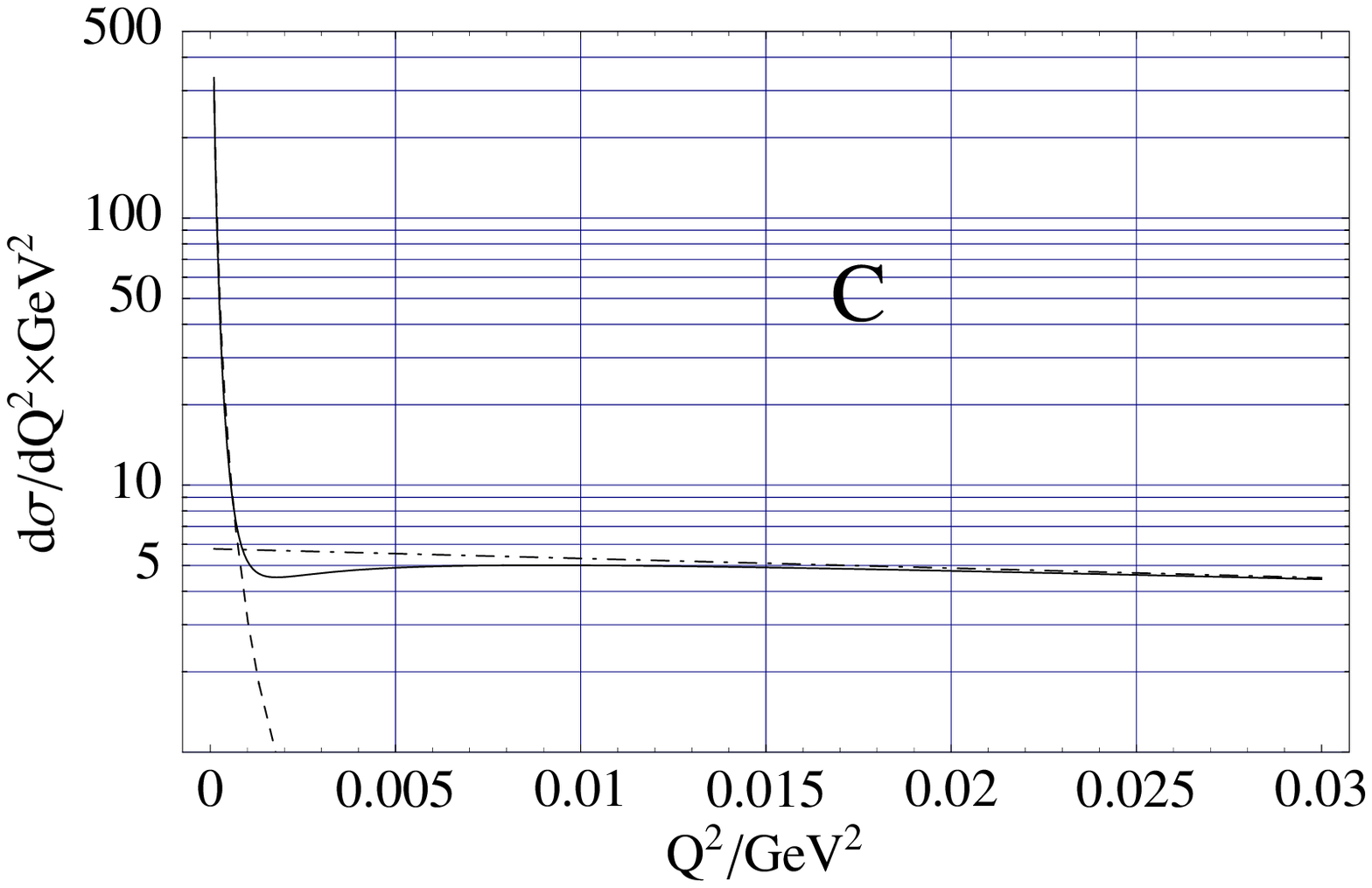}%
\vspace*{0mm}
\includegraphics[height=5cm]{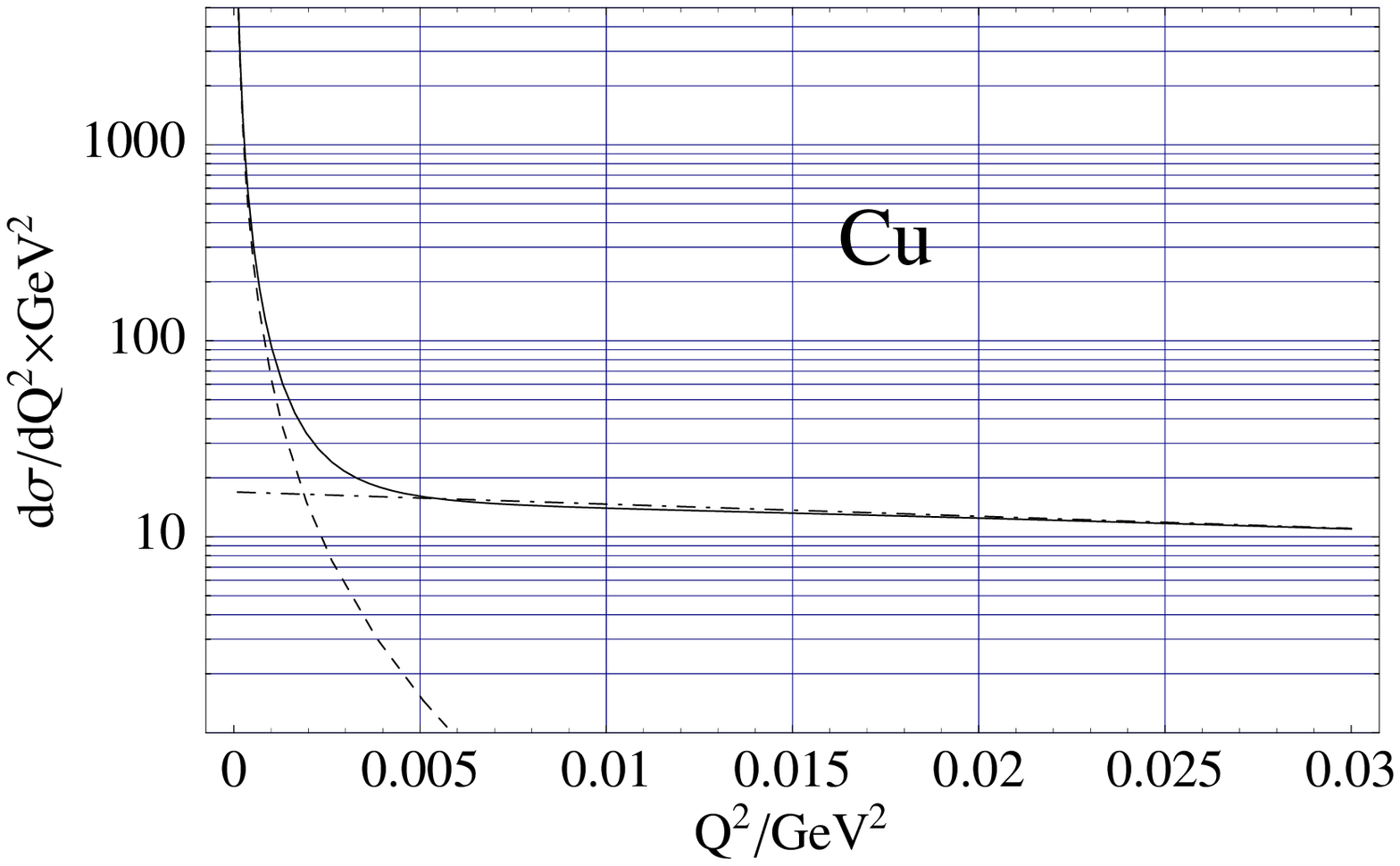} \\[2mm]
\includegraphics[height=5cm]{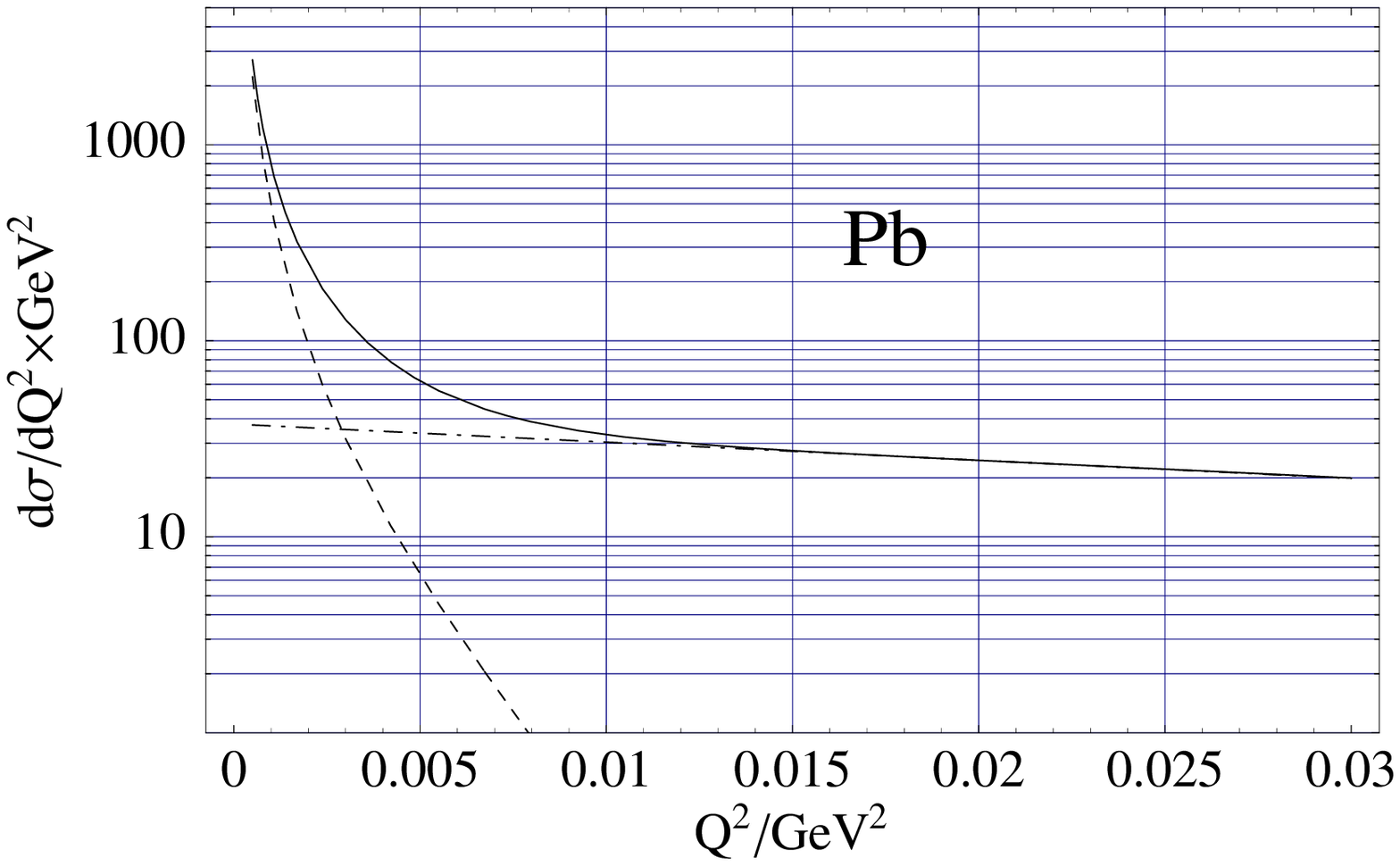}
\caption{The differential cross sections for the Primakoff scattering of a
  charged pion from different nuclei. The dashed curve signifies the Coulomb 
  part, the dash-dotted curve the strong part, and the solid curve the
  interference of Coulomb and strong amplitude.}
\label{fig:Coul-Strong-Interf}
\end{center}
\end{figure}

A useful simple model showing the interference between the Coulomb amplitude
and the diffraction amplitude is due to Locher \cite{Locher:1967so}. It has
been successfully used to describe the Coulomb-strong interference at high 
\cite{Jenni:1975ax} and at low energies \cite{Brueckner:1985hu}. 
Figure~\ref{fig:Coul-Strong-Interf} shows the results of the Locher model 
applied to the kinematical situation of the COMPASS experiment.

The free parameters  of the diffractive  part are the absolute height and 
the so called slope parameter $b$. They scale with the mass number
$A$ as $d \sigma / d \Omega|_{\text{diffractive}} \propto A$ and $b \propto
A^{1/3}$. This as well as the known electric form factor of the proton has 
been included in the calculation producing Fig.~\ref{fig:Coul-Strong-Interf}. 
The only parameter adjusted ``by hand'' is the relative phase between the 
Coulomb and the diffractive amplitude. Once adjusted to reproduce the marked 
interference minimum at $Q^2 \approx 3 \times 10^{-3}$\,GeV$^2$ for light
nuclei, clearly visible in the Serpukov data \cite{Antipov:1983df} as well 
as in the COMPASS data for carbon, it was kept fixed. If one now goes from
light nuclei to heavy nuclei the Coulomb phase changes and the destructive
interference becomes constructive. This has the consequence that for Pb the 
total differential cross section contains a considerable contribution from
strong interaction down to $Q^2 \approx 0.001\,$GeV$^2$. Therefore, the cut 
at the maximal $Q^2_{\text{max}} < 0.001\,$GeV$^2$ was well considered in 
the Serpukov experiment and the final analysis of COMPASS should respect 
this point.   

In summary, the question of the polarisability of the charged pion is
unsettled. However, the majority of the experimental results point to a 
value deviating by a factor of two from ChPTh. Since this prediction is one 
of the most stringent of ChPTh, the final result of COMPASS will be very
significant indeed. The recent criticism of F\"aldt \cite{Faldt:2006qh} 
performing a calculation of the Primakoff scattering using Glauber theory is 
fortunately not valid for the situation at Serpukov and COMPASS since the cuts 
in both experiments exclude the vector meson contributions this calculation 
includes.  

\section{Double polarised $\eta$ production with MAMI C}

As the last example of a significant observable we want to present a new
determination of the relative phase between the $S_{11}(1535)$ and the
$D_{13}(1520)$ excitations of the nucleon. This observable offers a sensitive
check whether the hypothesis that the nucleon excitation spectrum is composed 
of many broad overlapping Breit-Wigner resonances sitting on a non resonant 
``background'' is correct.

In the constituent quark model these resonances are explained by three quarks
bound in a QCD inspired, ad hoc confining potential. The excited state are by
construction narrow and a non-resonant background can not emerge from such a
model. The next best modelling is in the introduction of a coupling to the
meson decay channels. If one performs a full coupled channel calculation one
can model the width of the resonances and, it is believed, also of the
background. 

However, generally a Hamiltonian in quantum mechanics has a spectrum 
consisting of discrete states and a continuum both reflecting the dynamics 
of the considered system. The decay of the discrete states is mostly described 
by some imaginary potential, somewhat outside the laws of quantum mechanics, 
or as just mentioned, by coupled channel calculations. This means that it is 
rather unclear whether the continuum spectrum of the Hamiltonian is modelled 
or not by these approaches. It may very well be that some dynamics, actually 
present in nature, is missed because it is not included in the model
Hamiltonian. 

This problem is well known from nuclear physics and has limited
the ability  to explain the excitation spectrum of nuclei at high energies 
in the region of the ``giant resonances''. Here it is quite clear that much
strength is not due to giant resonances and the notion of missing resonances
has never been considered. Rather it came as a surprise that the strength of
some states was so much localised in energy in the many body system nucleus 
\cite{Mottelson:1976ac}. In the nucleus one distinguishes between the 
``spreading  width'' representing the spectrum of the continuum and the 
``decay width'' representing the coupling to some energetically open decay 
channels. The situation in nucleons is even more problematic since its
constituents, the  constituent quarks, are only very approximate effective
degrees of freedom. The excitation energy is of the same size as the
constituent mass and one can not expect that their mass and dynamics will 
not change with the excitation energy. For a more detailed account of these 
aspects see ref.~\cite{Drechsel:2007aa}.

The only way to make progress here is to go beyond the fitting of the
excitation spectrum with overlapping resonances. One has to perform 
experiments sensitive to all amplitudes including their phases in a 
given excitation energy region. This means the need to study interference  
terms as they are accessible with spin degrees of freedom. Through the 
coming into operation of the 1.5~GeV stage of the Mainz Microtron MAMI 
(MAMI C) beginning of the year 2007, it became possible to study the 
double polarisation cross section for the $\eta$ electroproduction on 
the proton with high precision \cite{Merkel:2007ig}.            

MAMI C is a double sided harmonic microtron having all the attractive features
of a modern electron accelerator: dc operation with high intensity of more than
100\,$\mu$A current, a small phase space providing a clean beam with very
small diameter, and polarised electrons and photons. A more detailed
description can be found in ref.~\cite{Walcher:2005ki}. 

It is known that the $S_{11}(1535)$ has a branching ratio of 50\% into the
$\eta\,p$ decay cannel, whereas the $D_{13}(1520)$ with a branching ratio of 
0.06\% couples only very weakly to this channel. A group at the Bonn electron 
storage ring ELSA had found a very peculiar angular behaviour of the target
polarisation asymmetry at the energy of these overlapping resonances 
\cite{Bock:1998rk} which could not be explained by the phenomenological model 
of ref.~\cite{Tiator:1999gr} without changing the relative phase of these two
resonances. This means that one gives up the assumption of two interfering
resonances described by Breit-Wigner shapes. A particularly intriguing
explanation supporting the concerns of the introduction to this section, may 
be the dynamical model of ref.~\cite{Kaiser:1996js}. In this model the
$S_{11}$ partial wave is described without the assumption of a resonance in 
the frame work of chiral dynamics with coupled channels.

The experiment at MAMI C measured the recoil polarisation of the 
$p(\vec{e},e'\vec{p})\eta$ reaction in the mass region of the $S_{11}(1535)$ 
and $D_{13}(1520)$ partial waves. The double polarisations $P^h_{x'}$ and
$P^h_{z'}$ agree well with the eta-MAID \cite{Chiang:2001as} assuming normal 
resonance behaviour. However, for the $P_{y'}$ single polarisation a three
sigma deviation from  the eta-MAID prediction was found \cite{Merkel:2007ig}. 
The single polarisation in the electro production $P_{y'}$ contains the same 
multipole combination $\mathcal{I}m \{E_{0+}^*(E_{2-}+M_{2-})\}$ as the 
target polarisation asymmetry in the photo production experiment at ELSA 
\cite{Bock:1998rk}. Consequently, as for the ELSA experiment, eta-MAID can 
only be made to agree by adjusting the relative phase between the
$S_{11}(1535)$ and $D_{13}(1520)$ states.  More such examples should be
studied  in order to learn better how to distinguish between resonances 
and background.
 
\section{Conclusions}
With three example we have shown that by selecting significant
observables one can trace the best effective degrees of freedom. The
deviations between theory and experiment indicated may force to go beyond 
the much used constituent quarks and Goldstone pions. Other effective
degrees of freedom may be the observed real ($K,~\rho,~\omega$ or virtual 
($\sigma$,~di-quarks) states. The aspect making observables significant is 
a deep theoretical understanding and the existence of reliable calculations 
based on this understanding on the one side. On the other side, the
experiments have to provide a matching accuracy to distinguish between
different models. Thriving for observables fulfilling these requirements is 
more significant than the investigation of more and more reactions with 
limited precision.

\end{document}